\documentclass[prd,showpacs,twocolumn,amsmath,amssymb]{revtex4}
\usepackage{graphicx,color,slashbox}
\begin{document}


\title{Bottom Baryons}

\author{Xiang Liu$^1$}
\email{xiangliu@pku.edu.cn}

\author{Hua-Xing Chen$^{1,2}$}
%
\author{Yan-Rui Liu$^{3}$}

\author{Atsushi Hosaka$^2$}
%
\author{Shi-Lin Zhu$^1$}
\email{zhusl@phy.pku.edu.cn}

\affiliation{$^1$School of Physics, Peking University, Beijing
100871, China \\ $^2$Research Center for Nuclear Physics, Osaka
University, Ibaraki 567--0047, Japan\\
$^3$ Institute of High Energy Physics, P.O. Box 918-4, Beijing
100049, China}

\date{\today}
\begin{abstract}

Recently CDF and D0 collaborations observed several bottom
baryons. In this work we perform a systematic study of the masses
of bottom baryons up to $1/m_Q$ in the framework of heavy quark
effective field theory (HQET) using the QCD sum rule approach. The
extracted chromo-magnetic splitting between the bottom baryon
heavy doublet agrees well with the experimental data.

\end{abstract}


\maketitle

\section{Introduction}\label{sec1}

Recently CDF Collaboration observed four bottom baryons
$\Sigma_{b}^{\pm}$ and $\Sigma_{b}^{*\pm}$ ~\cite{CDF,CDF-1}. D0
Collaboration announced the observation of $\Xi_b$~\cite{D0-Xi},
which was confirmed by CDF collaboration
later~\cite{CDF-Xi,CDF-Xi-1}. Very recently, Babar Collaboration
reported the observation of $\Omega_c^*$ with the mass splitting
$m_{\Omega_c^*}-m_{\Omega_c}=70.8\pm1.0\pm1.1$ MeV
\cite{Omega-star}. We collect the masses of these recently
observed bottom baryons in Table ~\ref{bottom-baryons}.

\begin{table}
\caption{The masses of bottom baryons recently observed by CDF and
D0 collaborations.}
\begin{center}
\begin{tabular}{c||cc}
\hline& mass (MeV)& Experiment
\\\hline\hline
$\Sigma_{b}^+$&$5808^{+2.0}_{-2.3}({\rm
stat.})\pm1.7({\rm syst.})$& \\
$\Sigma_b^-$&$5816^{+1.0}_{-1.0}({\rm stat.})\pm1.7({\rm syst.})$&\\
$\Sigma_{b}^{*+}$&$5829^{+1.6}_{-1.8}({\rm stat.})\pm1.7({\rm syst.})$&\raisebox{2ex}{CDF\cite{CDF,CDF-1}}\\
$\Sigma_{b}^{*-}$&$5837^{+2.1}_{-1.9}({\rm
stat.})\pm1.7({\rm syst.})$&\\\hline &$5774\pm11({\rm stat.})\pm15({\rm syst.})$& D0 \cite{D0-Xi}\\
\raisebox{1ex}{ $\Xi_b^-$}&$5793\pm 2.5({\rm stat.})\pm 1.7({\rm
syst.})$& CDF\cite{CDF-Xi,CDF-Xi-1}\\\hline
\end{tabular}\label{bottom-baryons}
\end{center}
\end{table}

The heavy hadron containing a single heavy quark is particularly
interesting. The light degrees of freedom (quarks and gluons)
circle around the nearly static heavy quark. Such a system behaves
as the QCD analogue of the familiar hydrogen bounded by
electromagnetic interaction. The heavy quark expansion provides a
systematic tool for heavy hadrons. When the heavy quark mass
$m_Q\to \infty$, the angular momentum of the light degree of
freedom is a good quantum number. Therefore heavy hadrons form
doublets. For example, $\Omega_b$ and $\Omega^\ast_b$ will be
degenerate in the heavy quark limit. Their mass splitting is
caused by the chromo-magnetic interaction at the order ${\cal
O}(1/m_Q)$, which can be taken into account systematically in the
framework of heavy quark effective field theory (HQET).

In the past two decades, various phenomenological models have been
used to study heavy baryon masses
~\cite{Capstick:1986bm,Roncaglia:1995az,Jenkins:1996de,
Mathur:2002ce,Ebert:2005xj,sum-rule-heavy-1}. Capstick and Isgur
studied the heavy baryon system in a relativized quark potential
model \cite{Capstick:1986bm}. Roncaglia et al. predicted the
masses of baryons containing one or two heavy quarks using the
Feynman-Hellmann theorem and semiempirical mass formulas
\cite{Roncaglia:1995az}. Jenkins studied heavy baryon masses using
a combined expansion of $1/{m_Q}$ and $1/{N_c}$
\cite{Jenkins:1996de}. Mathur et al. predicted the masses of
charmed and bottom baryons from lattice QCD \cite{Mathur:2002ce}.
Ebert et al. calculated the masses of heavy baryons with the
light-diquark approximation \cite{Ebert:2005xj}. Using the
relativistic Faddeev approach, Gerasyuta and Ivanov calculated the
masses of the S-wave charmed baryons \cite{GI}. Later, Gerasyuta
and Matskevich studied the charmed ${\bf (70,1^-)}$ baryon
multiplet using the same approach \cite{GM}. Stimulated by recent
experimental progress, there have been several theoretical papers
on the the masses of $\Sigma_{b}$, $\Sigma_{b}^{*}$ and $\Xi_b$
using the hyperfine interaction in the quark model
\cite{rosner,lipkin,rosner-Xi,Rosner-Lipkin,Spin-KL}. Recently the
strong decays of heavy baryons were investigated systematically
using $^3P_0$ model in Ref. \cite{ChengChongPRD2007}.

QCD sum rule (QSR) is a useful non-perturbative method in hadron
physics \cite{sum rule}, which has been applied to study heavy
baryon masses previously
~\cite{sum-rule-heavy-1,sum-rule-heavy-2,Navarra:1998vi,shuryak,grozin,sum-rule-heavy-3,dai,zhu,csh,hmq,Wang:2007sq,Wang:2002ts,Nielsen}.
The mass sum rules of $\Lambda_{c,b}$ and $\Sigma_{c,b}$ were
obtained in full QCD in Refs.
\cite{sum-rule-heavy-1,sum-rule-heavy-2,Navarra:1998vi}. The mass
sum rules of $\Sigma_Q$ and $\Lambda_Q$ in the leading order of
the heavy quark effective theory (HQET) have been discussed in
Refs. \cite{shuryak,grozin,sum-rule-heavy-3}. Dai et al.
calculated the $1/m_Q$ correction to the mass sum rules of
$\Lambda_Q$ and $\Sigma_Q^{(*)}$ in HQET \cite{dai}. Later the
mass sum rules of $\Lambda_Q$ and $\Sigma_Q^{(*)}$ were reanalyzed
in Ref. \cite{Wang:2002ts}. The mass sum rules of orbitally
excited heavy baryons in the leading order of HQET were discussed
in Refs. \cite{zhu,csh} while the $1/m_Q$ correction was
considered in Ref. \cite{hmq}. Recently Wang studied the mass sum
rule of $\Omega_{c,b}^{*}$ \cite{Wang:2007sq} while Dur\~{a}es and
Nielsen studied the mass sum rule of $\Xi_{c,b}$ using full QCD
Lagrangian \cite{Nielsen}.

In order to extract the chromo-magnetic splitting between the bottom
baryon doublets reliably, we derive the mass sum rules up to the
order of $1/m_Q$ in the heavy quark effective field theory in this
work. We perform a systematic study of the masses of $\Xi_b$,
$\Xi_{b}'$, $\Xi_b^*$, $\Omega_b$ and $\Omega_b^{*}$ through the
inclusion of the strange quark mass correction. The resulting
chromo-magnetic mass splitting agrees well with the available
experimental data. As a cross-check, we reproduce the mass sum rules
of $\Lambda_b$, $\Sigma_b$ and $\Sigma^\ast_b$ which have been
derived in literature previously. As a byproduct, we extend the same
formalism to the case of charmed baryons while keeping in mind that
the heavy quark expansion does not work well for the charmed
hadrons.

This paper is organized as follows. We present the formulation of
the leading order QCD sum rules in HQET for bottom baryons in
Section~\ref{sec2}. The following section is about the $1/m_Q$
correction. The numerical analysis and a short discussion are
presented in Section ~\ref{sec4}.

\section{QCD sum rules for heavy baryons}\label{sec2}

\begin{figure}[htb]
\begin{center}
\scalebox{0.9}{\includegraphics{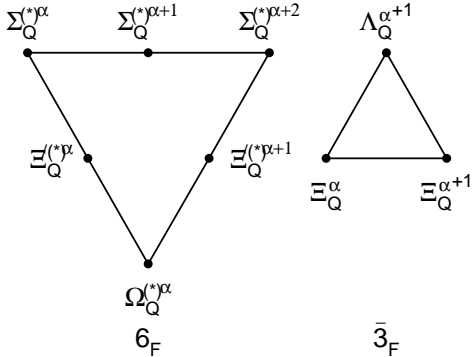}}
\end{center}
\caption{The SU(3) flavor multiplets of heavy baryons. Here
$\alpha$, $\alpha+1$, $\alpha+2$ denote the charges of heavy
baryons.\label{baryon}}
\end{figure}

We first introduce our notations for the heavy baryons. Inside a
heavy baryon there are one heavy quark and two light quarks ($u$,
$d$ or $s$). It belongs to either the symmetric $\mathbf{6_F}$ or
antisymmetric $\mathbf{\bar{3}_F}$ flavor representation (see
Fig.~\ref{baryon}). For the S-wave heavy baryons, the total
flavor-spin wave function of the two light quarks must be
symmetric since their color wave function is antisymmetric. Hence
the spin of the two light quarks is either $S=1$ for
$\mathbf{6_F}$ or $S=0$ for $\mathbf{\bar{3}_F}$. The angular
momentum and parity of the S-wave heavy baryons are
$J^{P}=\frac{1}{2}^+$ or $\frac{3}{2}^{+}$ for $\mathbf{6_F}$ and
$J^{P}=\frac{1}{2}^{+}$ for $\mathbf{\bar{3}_F}$. The names of
S-wave heavy baryons are listed in Fig.~\ref{baryon}, where we use
$*$ to denote $\frac{3}{2}^{+}$ baryons and the $\prime$ to denote
the $J^P=\frac{1}{2}^{+}$ baryons in the $\mathbf{6_F}$
representation. In this work, we use $B$ to denote the heavy
baryons with $\frac{1}{2}^+$ in $\mathbf{\bar{3}_{F}}$ and $B'$
and $B^*$ to denote those states with $\frac{1}{2}^+$ and
$\frac{3}{2}^+$ in $\mathbf{6_{F}}$.

We will study heavy baryon masses in HQET using QCD sum rule
approach. HQET plays an important role in the investigation of the
heavy hadron properties \cite{neubert}. In the limit of $m_Q\to
\infty$, the heavy quark field $Q(x)$ in full QCD can be
decomposed into its small and large components
\begin{eqnarray}
Q(x)=e^{-im_Q v\cdot x}[H_v(x)+h_v(x)],
\end{eqnarray}
where $v^{\mu}$ is the velocity of the heavy baryon. Accordingly
the heavy quark field $h_v(x)$ reads
\begin{eqnarray}
h_v(x)&=&e^{im_Q v\cdot x}\frac{1+v\!\!\!\slash}{2}Q(x),\\
 H_{v}(x)&=&e^{im_Q v\cdot x}\frac{1-v\!\!\!\slash}{2}Q(x).
\end{eqnarray}
The Lagrangian in HQET reads
\begin{eqnarray}
\mathcal{L}_{HQET}&=&\bar{h}_{v}iv\cdot D h_v
+\frac{1}{2m_Q}\bar{h}_{v}(iD_{\perp})^2
h_v\nonumber\\&&-C_{mag}\frac{g}{4m_Q}\bar{h}_v
\sigma_{\mu\nu}G^{\mu\nu}h_v.\label{heavy}
\end{eqnarray}
The second and third term in the above Lagrangian corresponds to
the kinetic and chromo-magnetic corrections at the order of
$1/m_Q$. Here $D^{\mu}_{\perp}=D^{\mu}-v^{\mu}v\cdot D$ and
$D^{\mu}=\partial^{\mu}+igA^{\mu}$. $C_{mag}(\mu)$ is
renormalization coefficient
$C_{mag}(\mu)=(\alpha_s(m_Q)/\alpha_s(\mu))^{3/\beta_0}[1+\frac{13\alpha_s}{6\pi}]$,
where $\beta_0=11-2n_f/3$ and $n_f$ is the number of quark flavors
\cite{neubert}.

In order to derive the mass sum rules of $B$, $B'$ and $B^*$, we
use the following interpolating currents for the heavy baryons
with $J^{P}=\frac{1}{2}^{+}$ in $\mathbf{6_F}$,
\begin{eqnarray}
J_{B'}(x)&=&\epsilon_{abc}[q_{1}^{aT}(x)C\gamma_{\mu}q_{2}^{b}(x)]\gamma_{t}^{\mu}\gamma_{5}h_{v}^{c}(x),\\
\bar{J}_{B'}(x)&=&-\epsilon_{abc}\bar{h}_{v}^{c}(x)\gamma_{5}\gamma_{t}^{\mu}[\bar{q}_{2}^{b}(x)\gamma_{\mu}C\bar{q}_{1}^{aT}(x)].
\end{eqnarray}
For the heavy baryons with $J^{P}=\frac{3}{2}^{+}$ in
$\mathbf{6_F}$,
\begin{eqnarray}
J^{\mu}_{B^*}(x)&=&\epsilon_{abc}[q_{1}^{aT}(x)C\gamma_{\nu}q_{2}^{b}(x)]\nonumber\\&&\times\Big(-g_{t}^{\mu\nu}+\frac{1}{3}
\gamma^{\mu}_{t}\gamma_{t}^{\nu}\Big)h_{v}^{c}(x),\\
 \bar{J}^{\mu}_{B^*}(x)&=&
\epsilon_{abc}\bar{h}_{v}^{c}(x)\Big(-g_{t}^{\mu\nu}+\frac{1}{3}
\gamma_{t}^{\nu}\gamma^{\mu}_{t}\Big)\nonumber\\&&\times[\bar{q}_{2}^{b}(x)\gamma_{\nu}C\bar{q}_{1}^{aT}(x)].
\end{eqnarray}
For the heavy baryons with $J^{P}=\frac{1}{2}^{+}$ in
$\mathbf{\bar{3}_F}$
\begin{eqnarray}
J_{B}(x)&=&\epsilon_{abc}[q_{1}^{aT}(x)C\gamma_{5}q_{2}^{b}(x)]h_{v}^{c}(x),\\
\bar{J}_{B}(x)&=&-\epsilon_{abc}\bar{h}_{v}^{c}(x)[\bar{q}_{2}^{b}(x)\gamma_{5}C\bar{q}_{1}^{aT}(x)].
\end{eqnarray}
Here $a$, $b$ and $c$ are color indices, $q_{i}(x)$ denotes up,
down and strange quark fields. $T$ is the transpose matrix and $C$
is the charge conjugate matrix.
$g_{t}^{\mu\nu}=g^{\mu\nu}-v^{\mu}v^{\nu}$,
$\gamma_{t}^{\mu}=\gamma^{\mu}-v\!\!\!\slash v^{\mu}$.

The overlapping amplitudes of the interpolating currents with $B$,
$B'$ and $B^*$ are defined as
\begin{eqnarray} \langle
0|J_{B}|B\rangle&=&f_{B}u_{B},\\
\langle 0|J_{B'}|B'\rangle&=&f_{B'}u_{B'},\\
\langle 0|J^{\mu}_{B^{*}}|B^*\rangle&=&\frac{1}{\sqrt{3}}f_{B^*}
u_{B^*} ^{\mu},
\end{eqnarray}
where $u^{\mu}_{B^{*}}$ is the Rarita-Schwinger spinor in HQET.
$f_{B'}=f_{B^{*}}$ due to heavy quark symmetry.

The binding energy ${\bar \Lambda}_i$ is defined as the mass
difference between the heavy baryon and heavy quark when $m_Q\to
\infty$. In order to extract ${\bar \Lambda}_i$, we consider the
following correlation function
\begin{eqnarray}
&&i\int {\rm{d}}^{4} x\; e^{iq\cdot x}\langle
0|T\{J_{B^{(')}}(x)\bar{J}_{B^{(')}}(0)\}|0\rangle =
\frac{1+v\!\!\!\slash}{2}\Pi_{B^{(')}}(\omega),\nonumber\\
\end{eqnarray}
with $\omega=v\cdot q$.

The dispersion relation for $\Pi(\omega)$ is
\begin{eqnarray}
\Pi(\omega)=\int\frac{\rho(\omega^\prime)}{\omega^\prime-\omega-i\epsilon}\;{\rm
d} \omega^\prime,
\end{eqnarray}
where $\rho(\omega)$ denotes the spectral density in the limit of
$m_{Q}\to \infty$. At the phenomenological level,
\begin{eqnarray}
\Pi(\omega)=\frac{f^{2}_{i}}{\bar{\Lambda}_{i}-\omega}+{\rm
continuum}.
\end{eqnarray}

Making the Borel transformation with variable $\omega$, we obtain
\begin{eqnarray}
f^{2}_{i}e^{-\bar{\Lambda}_{i}/T}=\int^{\omega_{0}}_{0}\rho(\omega)e^{-\omega/T}{\rm
d} \omega,
\end{eqnarray}
where we have invoked the quark-hadron duality assumption and
approximated the continuum above $\omega_0$ with the perturbative
contribution at the quark-gluon level. The mass sum rules of $B$,
$B'$ and $B^*$ are
\begin{eqnarray}
&&{f_{B}^2
e^{-\bar{\Lambda}_{B}/T}}=\nonumber\\&&\int^{\omega_{B}}_{0}\Big[\frac{\omega^{5}}{20\pi^4}-\frac{(m_{{q_{1}}}^2
+m_{{q_{2}}}^2-m_{{q_{1}}}m_{{q_{2}}}){\omega^{3}}}{4\pi^{4}}\nonumber\\&&
+ \frac{\langle g^2 GG \rangle \omega}{128 \pi^4}  +
\frac{m_{{q_{2}}}\langle\bar{{q_{2}}}{q_{2}}\rangle
+m_{{q_{1}}}\langle\bar{{q_{1}}}{q_{1}}\rangle }{4\pi^2
}\omega\nonumber\\&&- \frac{
2m_{{q_{2}}}\langle\bar{{q_{1}}}{q_{1}}\rangle+2m_{{q_{1}}}\langle\bar{{q_{2}}}{q_{2}}\rangle}{4\pi^2
}\Big] e^{-\omega/T}{\rm d}\omega \nonumber\\&&-
\frac{m_{{q_{1}}}\langle g_{c}\bar{{q_{2}}}\sigma G
{q_{2}}\rangle+m_{{q_{2}}}\langle g_{c}\bar{{q_{1}}}\sigma G
{q_{1}}\rangle}{32\pi^2}\nonumber\\&& +\frac{m_{{q_{1}}}\langle
g_{c}\bar{{q_{1}}}\sigma G {q_{1}}\rangle+m_{{q_{2}}}\langle
g_{c}\bar{{q_{2}}}\sigma G {q_{2}}\rangle}{12\cdot 32\pi^2} +
\frac{\langle\bar{{q_{1}}}{q_{1}}\rangle\langle\bar{{q_{2}}}{q_{2}}\rangle}{6}
\nonumber\\&&+ \frac{\langle\bar{{q_{1}}}{q_{1}}\rangle\langle
g_{c}\bar{{q_{2}}}\sigma G
{q_{2}}\rangle+\langle\bar{{q_{2}}}{q_{2}}\rangle\langle
g_{c}\bar{{q_{1}}}\sigma G {q_{1}}\rangle}{96T^2},
\end{eqnarray}
\begin{eqnarray}
&&f_{B'}^2
e^{-\bar{\Lambda}_{B'}/T}=\nonumber\\&&\int^{\omega_{B'}}_{0}\Big[\frac{3\omega^{5}}{20\pi^4}
+\frac{(3m_{q_{1}} m_{q_{2}}-3m_{q_{1}}^{2}-3m_{q_{2}}^2)
\omega^{3}}{4\pi^{4}}\nonumber\\&&-\frac{\langle g^2 GG \rangle
\omega}{128
\pi^4}-\frac{6m_{q_{1}}\langle\bar{{q_{2}}}{q_{2}}\rangle
+6m_{q_{2}}\langle\bar{{q_{1}}}{q_{1}}\rangle}{4\pi^{2}}\omega\nonumber\\&&+\frac{3m_{q_{1}}\langle\bar{{q_{1}}}{q_{1}}\rangle
+3m_{q_{2}}\langle\bar{{q_{2}}}{q_{2}}\rangle}{4\pi^{2}}\omega\Big]
e^{-\omega/T}{\rm
d}\omega\nonumber\\&&+\frac{\langle\bar{{q_{1}}}{q_{1}}\rangle\langle\bar{{q_{2}}}{q_{2}}\rangle}{2}
- {3m_{q_{1}} \langle g_c \bar {q_{2}} \sigma G {q_{2}}
\rangle+3m_{q_{2}} \langle g_c \bar {q_{1}} \sigma G {q_{1}} \rangle
\over 32\pi^2}\nonumber\\&&+ {5m_{q_{1}} \langle g_c \bar {q_{1}}
\sigma G {q_{1}} \rangle+5m_{q_{2}} \langle g_c \bar {q_{2}} \sigma
G {q_{2}} \rangle \over 128\pi^2}
 \nonumber\\&&+\frac{
\langle\bar{{q_{2}}}{q_{2}}\rangle\langle g_{c}\bar{{q_{1}}}\sigma
G{q_{1}}\rangle +\langle\bar{{q_{1}}}{q_{1}}\rangle\langle
g_{c}\bar{{q_{2}}}\sigma G{q_{2}}\rangle }{32T^2}.
\end{eqnarray}

The mass sum rule of $B^*$ is same as that of $B'$ at the leading
order of HQET. In the above equations, $\langle \bar{q_i}q_i
\rangle$ is the quark condensates, $\langle g^2 GG \rangle$ is the
gluon condensate and $\langle g\bar{q_i}\sigma Gq_i \rangle$ is the
quark-gluon mixed condensate. The above sum rules have been derived
in the massless light quark limit in Refs.
\cite{shuryak,grozin,sum-rule-heavy-3,dai}. Up and down quark mass
correction is tiny for heavy baryons $\Lambda_b$, $\Sigma_b$ and
$\Sigma_b^\ast$. In this work we have included the finite quark mass
correction which is important for heavy baryons $\Xi_b$, $\Xi_{b}'$,
$\Xi_b^*$, $\Omega_b$ and $\Omega_b^{*}$.

The binding energy $\bar{\Lambda}_{i}$ can be extracted using the
following formula
\begin{eqnarray}
\bar{\Lambda}_i&=&\frac{T^2 {d\mathbb{R}_{i}}}{\mathbb{R}_{i}{d T}},
\end{eqnarray}
where $\mathbb{R}_{i}$ denotes the right-hand part in the above sum
rules.

\section{The $1/m_Q$ correction}\label{sec3}

In order to calculate the ${1}/{m_{Q}}$ correction, we insert the
heavy baryon eigen-state of the Hamiltonian up to the order
$\mathcal{O}(1/m_Q)$ into the correlation function
\begin{eqnarray}
i\int d^4 x e^{iq\cdot x}\langle0| T[J_i(x) \bar J_i(0)]|0\rangle.
\end{eqnarray}
Its pole contribution is
\begin{eqnarray}
\Pi(\omega)&=&\frac{(f+\delta f)^2}{(\bar{\Lambda}+\delta
m)-\omega}\nonumber\\&=&\frac{f^2}{\bar{\Lambda}-\omega}-\frac{f^2\delta
m}{(\bar{\Lambda}-\omega)^2}+\frac{2f\delta
f}{\bar{\Lambda}-\omega},\label{deltam}
\end{eqnarray}
where both $\delta m$ and $\delta f$ are $\mathcal{O}(1/m_Q)$.

We consider the three-point correlation function
\begin{eqnarray}
&&\frac{1+v\!\!\!/}{2}\delta^{O}
\Pi(\omega,\omega')\nonumber\\&&=i^2\int d^4 z d^4 y e^{ip\cdot
z}e^{ip'\cdot
y}\langle0|T[J_{i}(z)O(x)\bar{J}(y)]|0\rangle,\label{three}
\end{eqnarray}
where operators $O=\mathcal{K}$ and $\mathcal{S}$ correspond to
the kinetic energy and chromo-magnetic interaction in Eq.
(\ref{heavy}). The double dispersion relation for
$\delta^{O}\Pi(\omega,\omega')$ reads
\begin{eqnarray}
\delta^{O}\Pi(\omega,\omega')=\int_{0}^{\infty}ds\int_{0}^{\infty}ds'
\frac{\rho^{O}(s,s')}{(s-\omega)(s'-\omega')}.\label{hah}
\end{eqnarray}
At the hadronic level,
\begin{eqnarray}
&&\delta^{\mathcal{K}}\Pi(\omega,\omega')
=\frac{f^2\mathcal{K}_i}{(\bar{\Lambda}-\omega)(\bar{\Lambda}-\omega')}
+\cdots,\label{K}\\
&&\delta^{\mathcal{S}}\Pi(\omega,\omega')
=\frac{f^2\mathcal{S}_i}{(\bar{\Lambda}-\omega)(\bar{\Lambda}-\omega')}
+\cdots \label{S}
\end{eqnarray}
with
\begin{eqnarray}
\mathcal{K}_i&=&\frac{1}{2m_Q}\langle B_i
|\bar{h}_{v}(iD_{\perp})^2
h_v| B_{i} \rangle,\\
\mathcal{S}_i&=&-\frac{1}{4m_Q}\langle B_i |\bar{h}_{v}g
\sigma_{\mu}G^{\mu\nu} h_v| B_{i} \rangle.
\end{eqnarray}
After setting $\omega=\omega'$ in Eqs. (\ref{K}) and (\ref{S}) and
comparing them with Eq. (\ref{deltam}), we can extract $\delta m$
\begin{eqnarray}
\delta m_i=-(\mathcal{K}_i+C_{mag}\mathcal{S}_i).
\end{eqnarray}
Here the renormalization coefficient $C_{mag}$ for bottom baryons
is $C_{mag}\approx 0.8$ \cite{zhu}.

\begin{figure}[htb]
\begin{center}
\scalebox{0.6}{\includegraphics{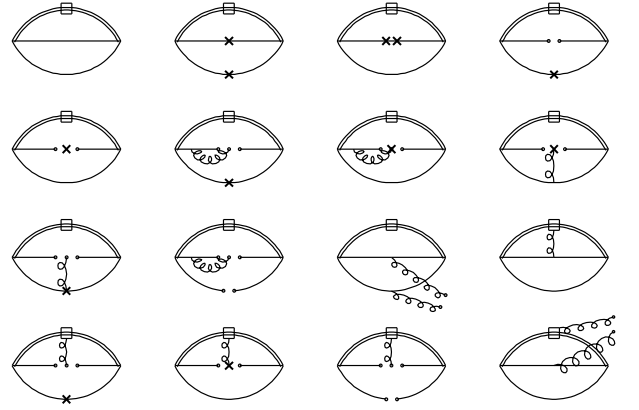}}
\end{center}
\caption{The diagrams for the $1/m_Q$ corrections. Here the
current quark mass correction is denoted by the cross. The first
eleven diagrams correspond to the kinetic corrections and the last
five diagrams are chromo-magnetic corrections. White squares
denote the operators of $1/m_Q$. \label{correction}}
\end{figure}

We calculate the diagrams listed in Fig. \ref{correction} to
derive $\delta^{O}\Pi(\omega,\omega')$. After invoking double
Borel transformation to Eq. \ref{hah}, we obtain the spectral
density $\rho^{O}(s,s')$. Then we redefine the integration
variable
\begin{equation}
s_{+}=\frac{s+s'}{2}\; ,
\end{equation}
\begin{equation}
s_{-}=\frac{s-s'}{2}\; .
\end{equation}
Now the integral in Eq. (\ref{hah}) is changed as
\begin{eqnarray}
\int_{0}^{\infty}ds\int_{0}^{\infty}ds'\ldots=2\int_{0}^{\infty}ds_+\int_{-s_+}^{+s_+}ds_-\ldots.
\end{eqnarray}
In the subtraction of the continuum contribution, quark hadron
duality is assumed for the integration variable $s_+$
\cite{variable}.

For $B(\frac{1}{2}^+)$ in $\mathbf{\bar{3}_F}$, the $1/m_Q$
correction comes from the kinetic term only.
\begin{eqnarray}
&&\mathcal{K}_{B}=\nonumber\\&&-\frac{e^{\bar{\Lambda}_B/T}}{m_Qf_B^2}
\Big\{\int^{\omega_{B}}_{0}\Big[\frac{54\omega^7}{7!\pi^4}
-\frac{9\omega^5}{5!\pi^4}(m_{q_1}^2+m_{q_2}^2-m_{q_1} m_{q_2})
\nonumber\\&&+\frac{3\langle g^2 GG \rangle \omega^3}{128\cdot 3!
\pi^4}+\frac{3\omega^3}{4\cdot
3!\pi^2}\Big(m_{q_1}\langle\bar{q_1}q_1\rangle+m_{q_2}\langle\bar{q_2}q_2\rangle\nonumber\\&&-
2m_{q_2}\langle\bar{q_1}q_1\rangle-2m_{q_1}\langle\bar{q_2}q_2\rangle\Big)\nonumber\\&&-\frac{3\omega}{128\pi^2}\Big(m_{q_1}\langle
g_c\bar{q_1}\sigma G q_1\rangle+m_{q_2}\langle g_c \bar{q_2}\sigma
G
q_2\rangle\Big)\nonumber\\&&+\frac{3\omega}{32\pi^2}\Big(m_{q_1}\langle
g_c\bar{q_2}\sigma G q_2\rangle+m_{q_2}\langle g_c \bar{q_1}\sigma
G q_1\rangle\Big)\Big]e^{-\omega/T}{\rm
d}\omega\nonumber\\
&&-\frac{1}{32}\Big[ \langle\bar{q_1}q_1\rangle\langle
g_{c}\bar{q_2}\sigma G
q_2\rangle+\langle\bar{q_2}q_2\rangle\langle g_{c}\bar{q_1}\sigma
G q_1\rangle
 \Big]\Big\},\\ \nonumber\\
&&\mathcal{S}_{B}=0.
\end{eqnarray}
Here $\mathcal{S}_{B}=0$ is consistent with the simple expectation
in the constituent quark model that the chromo-magnetic
interaction $\langle S_{Q}\cdot j_{l}\rangle=0$ since $j_l=0$ for
$B(\frac{1}{2}^+)$ in $\mathbf{\bar{3}_{F}}$.

For $B'(\frac{1}{2}^+)$ in $\mathbf{6_F}$, the $1/m_Q$ corrections
are
\begin{eqnarray}
&&\mathcal{K}_{B'}=\nonumber\\&&-\frac{e^{\bar{\Lambda}_{B'}/T}}{m_Qf_{B'}^2}\Big\{\int^{\omega_{B'}}_{0}
\Big[\frac{18\cdot11\omega^7}{7!\pi^4}
-\frac{9\omega^5}{5!\pi^4}(4m_{q_1}^2+4m_{q_2}^2\nonumber\\&&-3m_{q_1}
m_{q_2}) -\frac{\langle g^2 GG \rangle \omega^3}{128\cdot3!
\pi^4}+\frac{3\omega^3}{4\cdot
3!\pi^2}\Big(5m_{q_1}\langle\bar{q_1}q_1\rangle\nonumber\\
&&+5m_{q_2}\langle\bar{q_2}q_2\rangle-
6m_{q_2}\langle\bar{q_1}q_1\rangle-6m_{q_1}\langle\bar{q_2}q_2\rangle\Big)\nonumber\\&&{
+}\frac{11\omega}{128\cdot4\pi^2}\Big(m_{q_1}\langle
g_c\bar{q_1}\sigma G q_1\rangle+m_{q_2}\langle g_c \bar{q_2}\sigma G
q_2\rangle\Big)\Big]e^{-\omega/T}{\rm
d}\omega\nonumber\\
&&-\frac{3}{32}\Big[ \langle\bar{q_1}q_1\rangle\langle
g_{c}\bar{q_2}\sigma G
q_2\rangle+\langle\bar{q_2}q_2\rangle\langle g_{c}\bar{q_1}\sigma
G q_1\rangle
 \Big]\Big\}.
\end{eqnarray}
\begin{eqnarray}
&&\mathcal{S}_{B'}=\nonumber\\&&\frac{e^{\bar{\Lambda}_{B'}/T}}{m_Qf_{B'}^2}\Big\{\int^{\omega_{B'}}_{0}
\Big[\frac{2g_c^2\omega^7}{105\pi^6}+\frac{\langle g^2 GG \rangle
\omega^3}{16\cdot3!
\pi^4}\nonumber\\&&-\frac{\omega}{32\pi^2}\Big(m_{q_1}\langle
g_c\bar{q_1}\sigma G q_1\rangle+m_{q_2}\langle g_c \bar{q_2}\sigma
G q_2\rangle\nonumber\\&&-2m_{q_2}\langle g_c\bar{q_1}\sigma G
q_1\rangle-2m_{q_1}\langle g_c \bar{q_2}\sigma G
q_2\rangle\Big)\Big]e^{-\omega/T}{\rm
d}\omega\nonumber\\&&-\frac{1}{48}\Big[
\langle\bar{q_1}q_1\rangle\langle g_{c}\bar{q_2}\sigma G
q_2\rangle+\langle\bar{q_2}q_2\rangle\langle g_{c}\bar{q_1}\sigma
G q_1\rangle \Big]\Big\}.
\end{eqnarray}
Through explicit calculation, we obtain
\begin{equation}
\mathcal{K}_{B^*}=\mathcal{K}_{B'}\; ,
\end{equation}
\begin{equation}
\mathcal{S}_{B^*}=-{\mathcal{S}_{B'}/ 2}\; ,
\end{equation}
\begin{equation}
m_{B^*}-m_{B'}=\frac{3}{2}S_{B'},\label{split}
\end{equation}
which are consistent with the heavy quark symmetry.

\section{Results and discussion}\label{sec4}

In our numerical analysis, we use \cite{mass
splitting-1,Gasser:1984gg,Gimenez:2005nt,Jamin:2002ev,Ioffe:2002be,Ovchinnikov:1988gk,PDG}:
\begin{eqnarray*}
\nonumber &&\langle \bar q q \rangle = -(0.240 \;\mbox{ GeV})^3\, ,
\\ && \langle\bar ss\rangle=(0.8\pm 0.1)\times\langle \bar q q \rangle\, ,
\\ \nonumber && \langle g_s \bar q \sigma G q \rangle = - M_0^2
\times \langle \bar q q \rangle \, ,
\\ \nonumber && M_0^2 = ( 0.8 \pm 0.2 )\; \mbox{GeV}^2\, ,
\\
\nonumber &&\langle g_s^2 G^2 \rangle = (0.48 \pm 0.14) \mbox{
GeV}^4\, ,
\\ \nonumber && m_u =  m_d = 5.3 \;\mbox{MeV},~m_s=125 \mbox{MeV}\,
,
\\ \nonumber && m_c = 1.25 \pm 0.09 \;\mbox{GeV}, ~ m_b = 4.8 \;\mbox{GeV} \,
. \\ \nonumber && \alpha_s(m_c) = 0.328, ~ \alpha_s(m_b) = 0.189\, .
\end{eqnarray*}
The values of the $u,d,s$ and charm quark masses correspond to the
$\overline{MS}$ scheme at a scale $\mu\approx 2$ GeV and
$\mu=\overline{m}_c$ respectively \cite{PDG}. The $b$ quark mass
is obtained from the Upsilon $1S$ mass \cite{PDG,1S}.

Since the energy gap between the S-wave heavy baryons and their
radial/orbital excitations is around 500 MeV, the continuum
contribution can be subtracted quite cleanly. We require that the
high-order power corrections be less than 30\% of the perturbative
term to ensure the convergence of the operator product expansion.
This condition yields the minimum value for the working region of
the Borel parameter. In this work, we choose the working region as
$0.4<T<0.6$ GeV.

In Fig. \ref{a-1}-\ref{a-3}, we give the dependence of
$\bar{\Lambda}$, $\mathcal{K}_{i}$, $\mathcal{S}_{i}$ and mass
splitting  $m_{B_b^{*}}-m_{B_b'}$ on $T$ and $\omega_c$ for
$\Sigma_b$, $\Xi_b'$, $\Omega_b$. The variation of a sum rule with
both $T$ and $\omega_i$ contributes to the errors of the extracted
value, together with the truncation of the operator product
expansion and the uncertainty of vacuum condensate values. We
collect the extracted $\bar{\Lambda}$, $\mathcal{K}_{i}$,
$\mathcal{S}_{i}$ and mass splitting $m_{B_c^{*}}-m_{B_c'}$ in
Table~\ref{table2}.

The masses of bottom baryons from the present work are presented
in Table~\ref{table3}. It's well known that the heavy quark
expansion does not work very well for the charmed baryons since
the charm quark is not heavy enough to ensure the good convergence
of $1/m_Q$ expansion. For example, the chromo-magnetic splitting
between $\Omega_c^\ast$ and $\Omega_c$ from our work is around 133
MeV, which is much larger than the experimental value 67.4 MeV.
However, we still choose to present the masses of S-wave charmed
baryons also in Table~\ref{table3} simply for the sake of
comparison with experimental data.

\begin{widetext}
\begin{center}
\begin{table}[htb]
\caption{The central values in this table are extracted at $T=0.5$
GeV, $\omega_i=1.3$ GeV for $\Sigma_b^{(*)}$, $\omega_i=1.4$ GeV
for $\Xi_b^{'(*)}$, $\omega_i=1.55$ GeV for $\Omega_b^{(*)}$,
$\omega_i=1.1$ GeV for $\Lambda_b$ and $\omega_i=1.25$ GeV for
$\Xi_b$ (in MeV).}
\begin{center}
\begin{tabular}{c||ccc|ccccccc}
\hline & $\Sigma_b$ & $\Xi^{\prime}_b$ & $\Omega^0_b$ &
$\Lambda_b$ & $\Xi_b$
\\ \hline\hline 
  $\bar \Lambda$
&$950^{+78}_{-74}$&$1042^{+76}_{-74}$&$1169\pm{74}$&$773^{+68}_{-59}$&$908^{+72}_{-67}$
\\  $\delta m$   &$59^{+4}_{-2}$
&$60^{+6}_{-4}$&$67^{+7}_{-3}$&$65^{+2}_{-1}$&$72{\pm1}$
\\\hline\hline
mass splitting&$m_{\Sigma_b^{*}}-m_{\Sigma_b}$&$m_{\Xi_b^{*}}-m_{\Xi_b'}$&$m_{\Omega_b^{*}}-m_{\Omega_b}$&-&-\\
this work&$26{\pm1}$ &$26\pm1$&$28^{+8}_{-2}$&-&-\\
experiment \cite{CDF,CDF-1}&21&-&-&-&-&\\
 \hline
\end{tabular}\label{table2}
\end{center}
\end{table}

\begin{table}[htb]
\caption{Masses of the heavy baryons from the present work and
other approaches and the comparison with experimental data (in
MeV).}
\begin{tabular}{cc|ccccccc|c}
\hline Baryon & $I(J^P)$ & Ours &Ref.~\cite{Capstick:1986bm}
&Ref.~\cite{Roncaglia:1995az}
&Ref.~\cite{Jenkins:1996de}&Ref.~\cite{Mathur:2002ce} &
Ref.~\cite{Ebert:2005xj} & Ref.~\cite{Wang:2007sq,Wang:2002ts}
&EXP~\cite{PDG,CDF-1,D0-Xi,CDF-Xi,Omega-star}\\
\hline \hline
$\Sigma_c$&$1(\frac12^+)$&$2411^{+93}_{-81}$&2440&2453& &2452&2439&2470&2454.02(0.18)\\
$\Xi'_c$&$\frac12(\frac12^+)$&$2508^{+97}_{-91}$& &2580&2580.8&2599&2578& &2575.7(3.1)\\
$\Omega_c$ & $0(\frac12^+)$&$2657^{+102}_{-99}$& &2710& &2678&2698& &2697.5(2.6)\\
$\Sigma^*_c$&$1(\frac32^+)$&$2534^{+96}_{-81}$&2495&2520& &2538&2518&2590&2518.4(0.6)\\
$\Xi^*_c$&$\frac12(\frac32^+)$&$2634^{+102}_{-94}$& &2650& &2680&2654& &2646.6(1.4)\\
$\Omega^*_c$ & $0(\frac32^+)$&$2790^{+109}_{-105}$& &2770&2760.5&2752&2768&2790&$\sim 2768$\\
$\Lambda_c$&$0(\frac12^+)$&$2271^{+67}_{-49}$&2265&2285& &2290&2297& &2286.46(0.14)\\
$\Xi_c$&$\frac12(\frac12^+)$&$2432^{+79}_{-68}$& &2468& &2473&2481& &2467.9(0.4)\\
\hline \hline
$\Sigma_b$&$1(\frac12^+)$&$5809^{+82}_{-76}$&5795&5820&5824.2&5847&5805&5790&5808\\
$\Xi'_b$&$\frac12(\frac12^+)$&$5903^{+81}_{-79}$& &5950&5950.9&5936&5937& &\\
$\Omega_b$ & $0(\frac12^+)$&$6036\pm81$& &6060&6068.7&6040&6065& &\\
$\Sigma^*_b$&$1(\frac32^+)$&$5835^{+82}_{-77}$&5805&5850&5840.0&5871&5834&5820&5829\\
$\Xi^*_b$&$\frac12(\frac32^+)$&$5929^{+83}_{-79}$& &5980&5966.1&5959&5963& &\\
$\Omega^*_b$ & $0(\frac32^+)$&$6063^{+83}_{-82}$& &6090&6083.2&6060&6088&6000&\\
$\Lambda_b$&$0(\frac12^+)$&$5637^{+68}_{-56}$&5585&5620& &5672&5622& &5624(9)\\
$\Xi_b$&$\frac12(\frac12^+)$&$5780^{+73}_{-68}$& &5810&5805.7&5788&5812& &5774,5793\\
\hline
\end{tabular}\label{table3}
\end{table}
\end{center}
\end{widetext}

In our calculation, we adopt the phenomenological spectral
function by the classical and simple ansatz of a single resonance
pole plus the perturbative continuum. The systematic uncertainty
of hadron parameters obtained with such an approximation was
discussed recently in Ref. \cite{lms07}. We have not considered
the next-to-leading order $\alpha_s$ corrections, which may also
result in large contribution and uncertainty as indicated by the
study of the $\alpha_s$ corrections in the light-quark baryon
system in Ref. \cite{ops91}.

In short summary, inspired by recent experimental observation of
charmed and bottom baryons
\cite{CDF,CDF-1,D0-Xi,CDF-Xi,CDF-Xi-1,Omega-star}, we have
investigated the masses of heavy baryons systematically using the
QCD sum rule approach in HQET. The chromo-magnetic splitting of
the bottom baryon doublet from the present work agrees well with
the recent experimental data. Recently $\Xi_b^{(*)}$ was observed
by CDF collaboration \cite{CDF,CDF-1}. Our results are also
consistent with their experimental value. Our prediction of the
masses of $\Xi_b'$, $\Xi_b^*$, $\Omega_b$ and $\Omega_b^*$ can be
tested through the future discovery of these interesting states at
Tevatron at Fermi Lab.

\vfill
\section*{Acknowledgments}
\vfill

 X.L. thanks W. Wei for useful discussion. This project was
supported by the National Natural Science Foundation of China under
Grants 10421503, 10625521 and 10705001, Key Grant Project of Chinese
Ministry of Education (No. 305001) and the China Postdoctoral
Science foundation (No. 20060400376). H.X.C. is grateful to the
Monkasho fellowship for supporting his stay at Research Center for
Nuclear Physics where this work is done. A.H. is supported in part
by the Grant for Scientific Research ((C) No.19540297) from the
Ministry of Education, Culture, Science and Technology, Japan.

\begin{widetext}

\begin{center}
\begin{figure}[htb]
\begin{tabular}{cc}
\scalebox{0.8}{\includegraphics{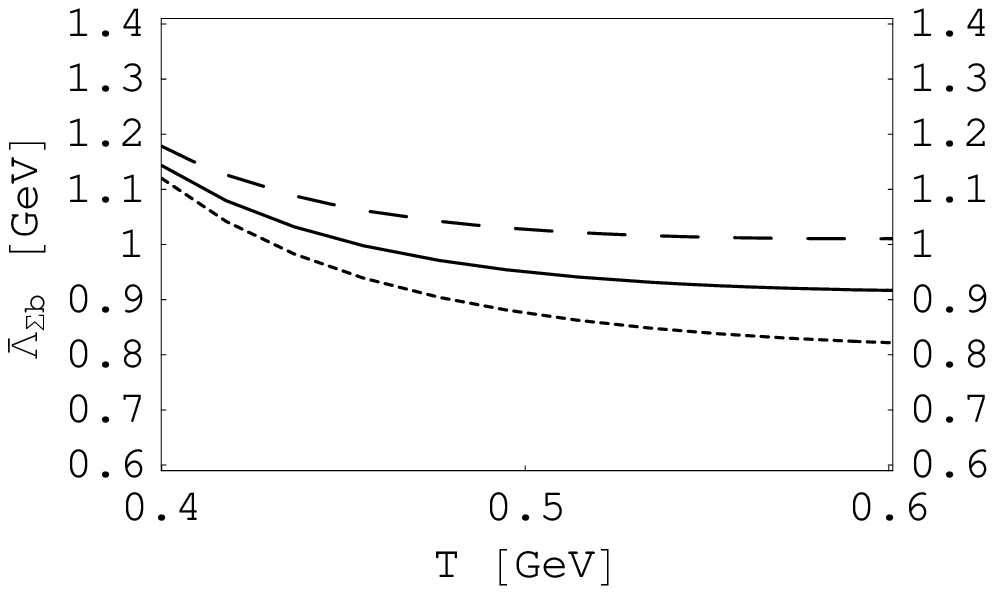}}&\scalebox{0.8}{\includegraphics{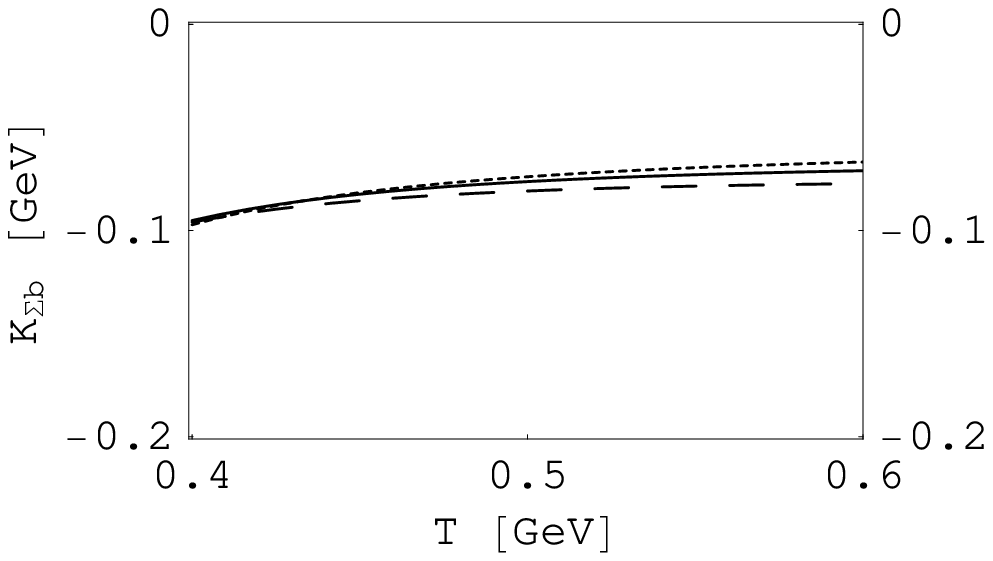}}\\
\scalebox{0.8}{\includegraphics{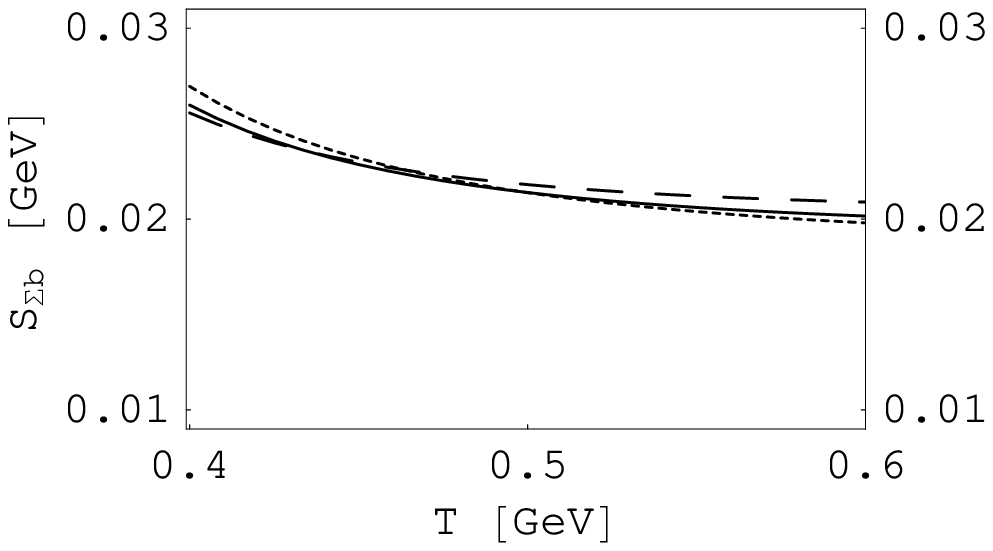}}&\scalebox{0.8}{\includegraphics{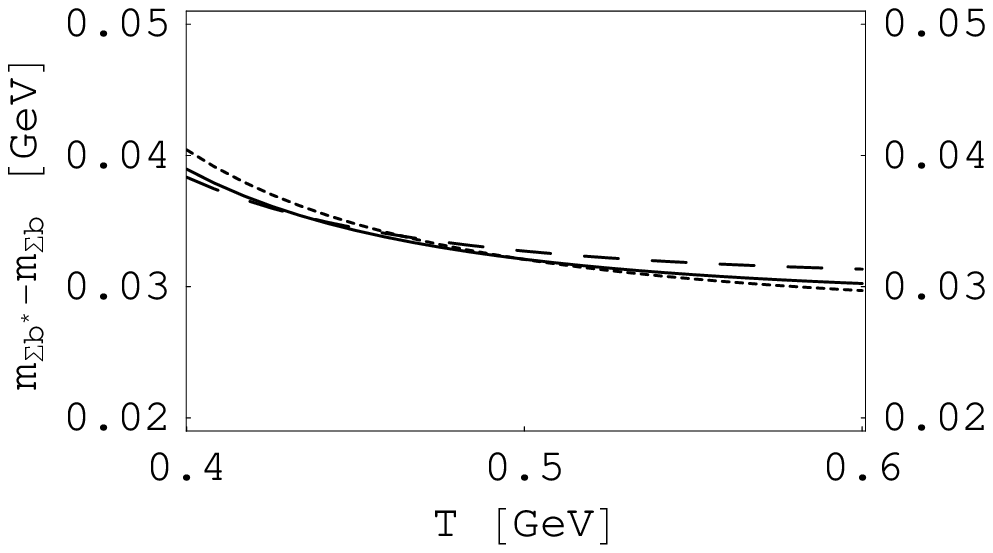}}\\
\end{tabular}
\caption{The dependences of $\bar{\Lambda}_{\Sigma_b}$,
$\mathcal{K}_{\Sigma_b}$, $\mathcal{S}_{\Sigma_b}$, and the mass
splitting $m_{\Sigma_b^{*}}-m_{\Sigma_b}$ on $T$. Here the dotted,
solid and dashed line corresponds to the threshold value
$\omega_{\Sigma_b}=1.2,1.3,1.4$ GeV respectively. }\label{a-1}
\end{figure}\end{center}

\begin{center}
\begin{figure}[htb]
\begin{tabular}{cc}
\scalebox{0.8}{\includegraphics{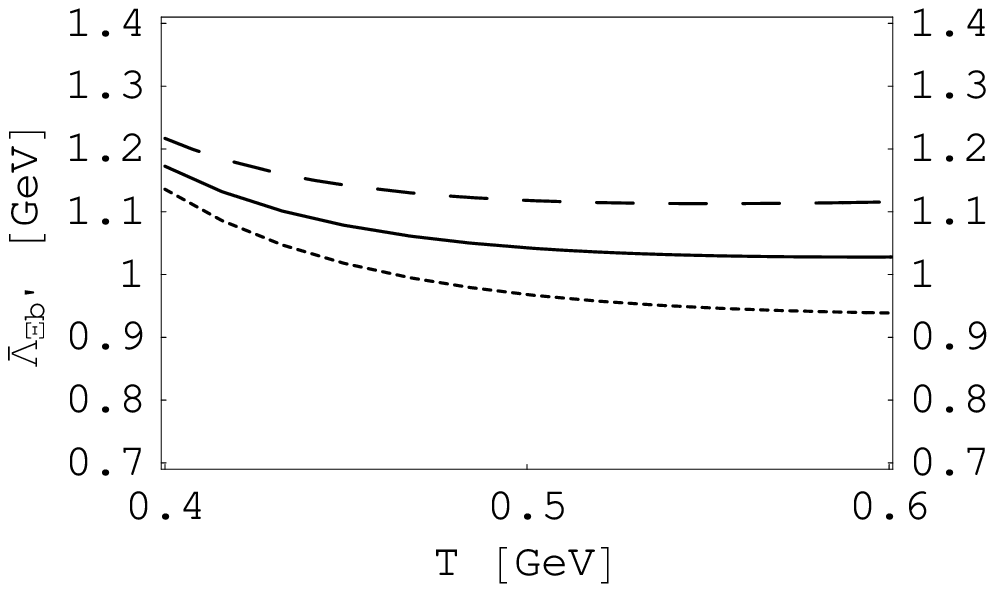}}&\scalebox{0.8}{\includegraphics{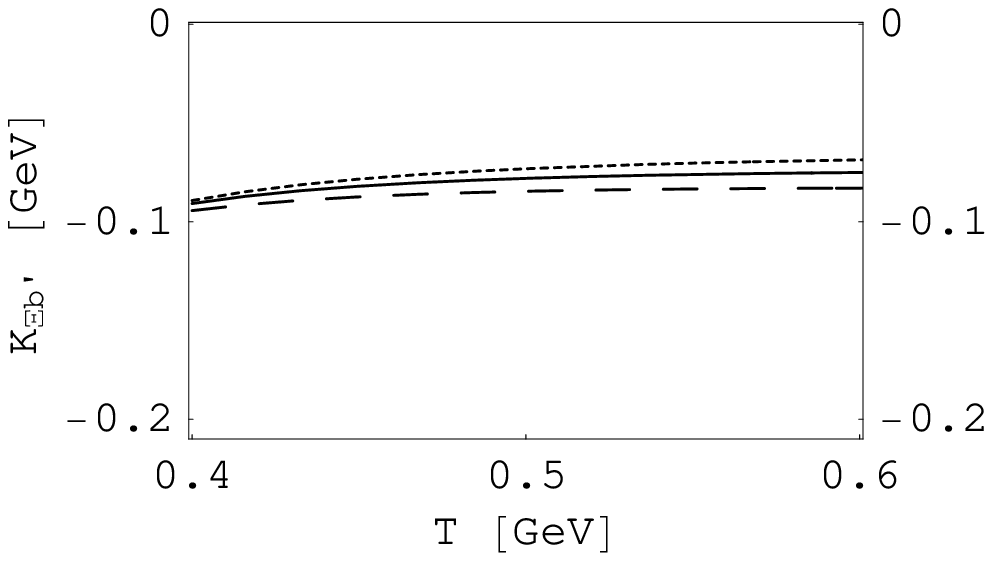}}\\
\scalebox{0.8}{\includegraphics{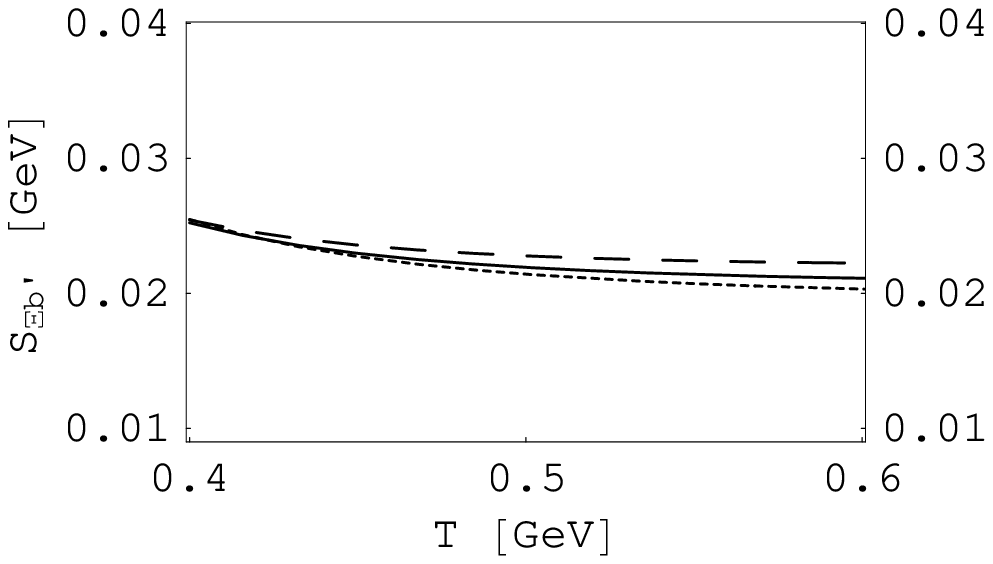}}&\scalebox{0.8}{\includegraphics{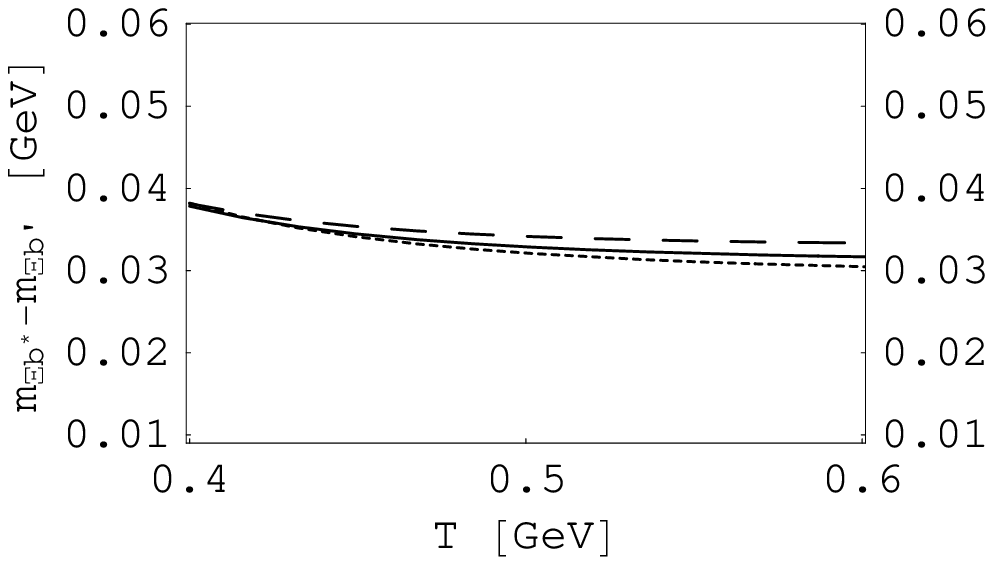}}\\
\end{tabular}
\caption{The dependences of $\bar{\Lambda}_{\Xi_b'}$,
$\mathcal{K}_{\Xi_b'}$, $\mathcal{S}_{\Xi_b'}$, and the mass
splitting $m_{\Xi_b^{*}}-m_{\Xi_b'}$ on $T$. The dotted, solid and
dashed line corresponds to $\omega_{\Xi_b'}=1.3,1.4,1.5$ GeV
respectively. }\label{a-2}
\end{figure}\end{center}

\begin{center}
\begin{figure}[htb]
\begin{tabular}{cc}
\scalebox{0.8}{\includegraphics{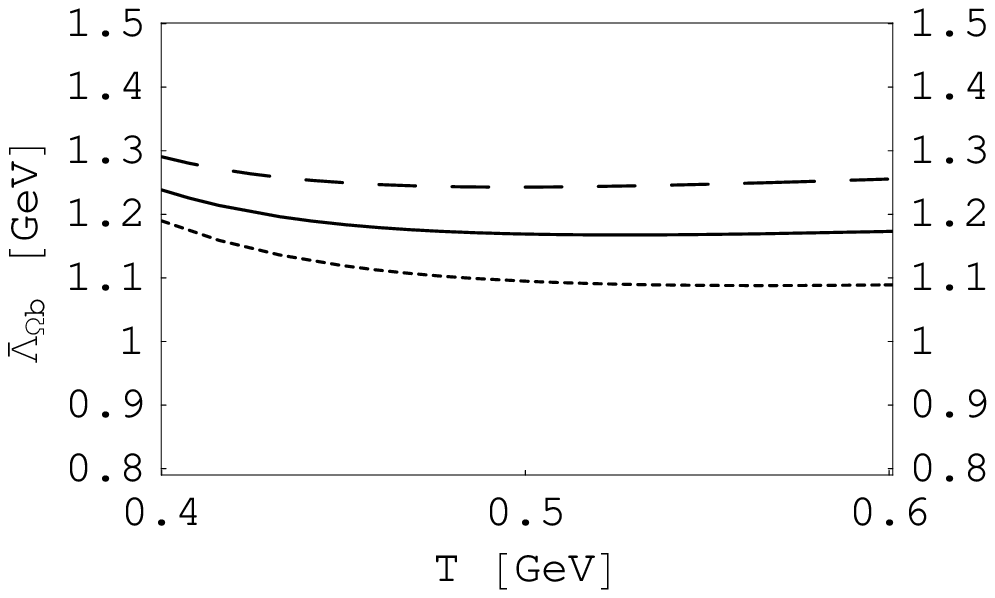}}&\scalebox{0.8}{\includegraphics{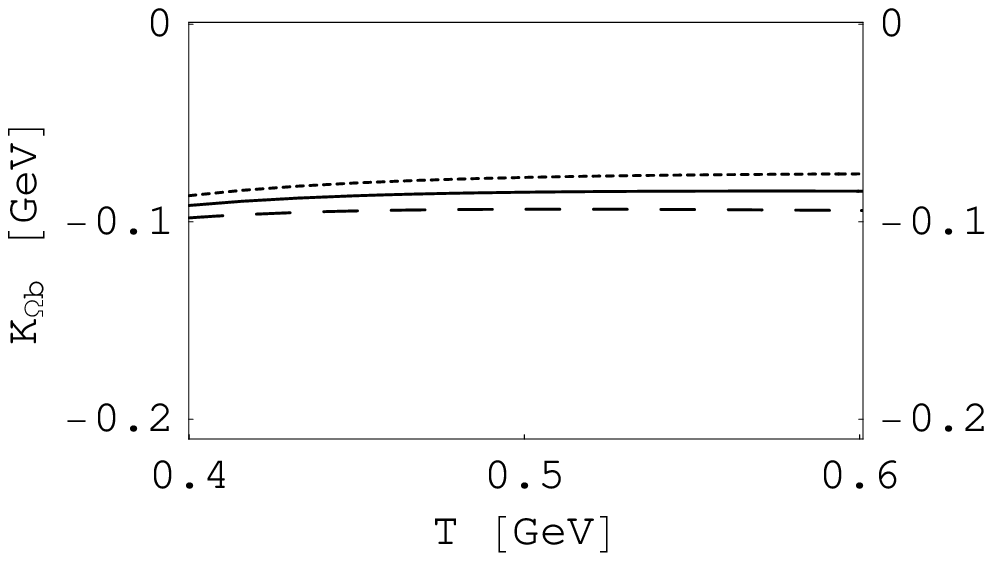}}\\
\scalebox{0.8}{\includegraphics{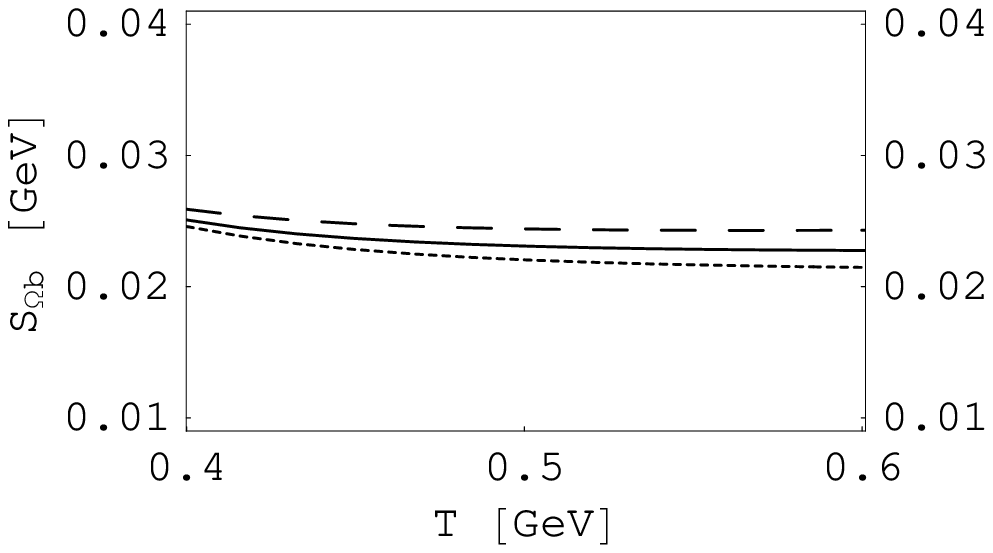}}&\scalebox{0.8}{\includegraphics{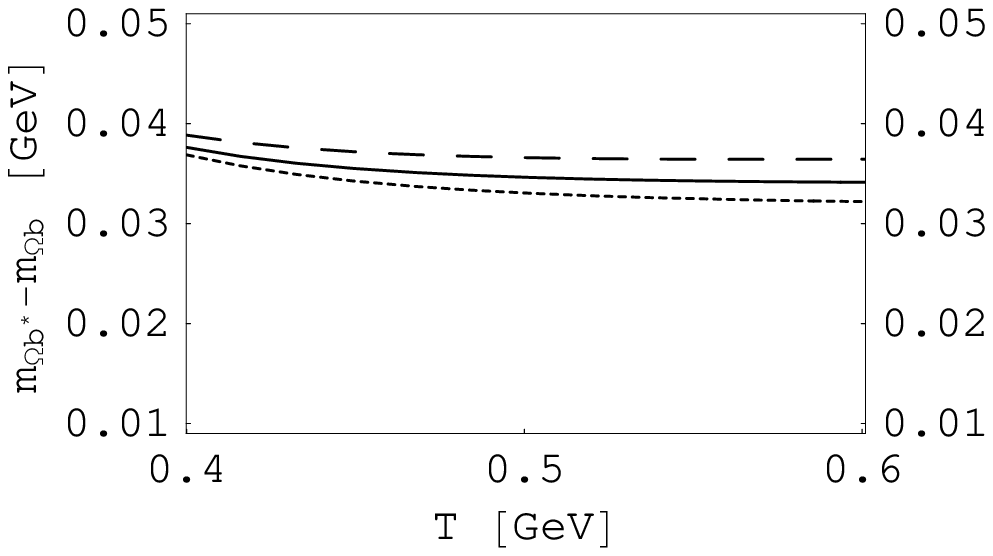}}\\
\end{tabular}
\caption{The dependences of $\bar{\Lambda}_{\Omega_b}$,
$\mathcal{K}_{\Omega_b}$, $\mathcal{S}_{\Omega_b}$, and the mass
splitting $m_{\Omega_b^{*}}-m_{\Omega_b}$ on $T$. The dotted,
solid and dashed line corresponds to
$\omega_{\Omega_b}=1.45,1.55,1.65$ GeV respectively. }\label{a-3}
\end{figure}\end{center}

\end{widetext}

\end{document}